\documentclass[aps,prb,twocolumn,amsmath,superscriptaddress]{revtex4}
\usepackage{graphicx}
\usepackage{color}

\begin{document}

\title{Dynamics of a Mn spin coupled to a single hole confined in a quantum dot}
\author{B. Varghese}
\affiliation{
CNRS, Institut N\'eel, F-38042 Grenoble, France.}
\affiliation{
Universit\'{e} Grenoble Alpes, Institut N\'eel, F-38042 Grenoble, France}

\author{H. Boukari}
\affiliation{
CNRS, Institut N\'eel, F-38042 Grenoble, France.}
\affiliation{
Universit\'{e} Grenoble Alpes, Institut N\'eel, F-38042 Grenoble, France}

\author{L. Besombes}
\email{lucien.besombes@grenoble.cnrs.fr}
\affiliation{
CNRS, Institut N\'eel, F-38042 Grenoble, France.}
\affiliation{
Universit\'{e} Grenoble Alpes, Institut N\'eel, F-38042 Grenoble, France}
\date{\today}

\begin{abstract}

Using the emission of the positively charged exciton as a probe, we analyze the dynamics of the optical pumping and the dynamics of the relaxation of a Mn spin exchange-coupled with a confined hole spin in a II-VI semiconductor quantum dot. The hole-Mn spin can be efficiently initialized in a few tens of $ns$ under optical injection of spin polarized carriers. We show that this optical pumping process and its dynamics are controlled by electron-Mn flip-flops within the positively charged exciton-Mn complex. The pumping mechanism and its magnetic field dependence are theoretically described by a model including the dynamics of the electron-Mn complex in the excited state and the dynamics of the hole-Mn complex in the ground state of the positively charged quantum dot. We measure at zero magnetic field a spin relaxation time of the hole-Mn spin in the $\mu s$ range or shorter. This hole-Mn spin relaxation is induced by the presence of valence band mixing in self-assembled quantum dots.

\end{abstract}

\maketitle

\section{Introduction}

In ferromagnets currently used in data storage devices, the magnetic anisotropy determines the stability of the orientation of the magnetization at a given temperature and eventually in the presence of an external applied magnetic field. This magnetic anisotropy is an intrinsic property of the magnetic material which depends on the local environment of the magnetic atoms. In the last ten years, experimental breakthroughs have laid the foundations for atomic scale data storage, showing the capability to read and manipulate the spin of a single magnetic atom. In particular, it has been shown that the isotropic magnetic moment of a free atom can develop a magnetic anisotropy energy when it interacts with the ordered surface of a metal \cite{Gambardella2003,Oberg2014}. An individual magnetic atom can also present a large magnetic anisotropy when interacting with ligands in a molecular magnet \cite{Bogani2008}. In semiconductors, it has been shown that the optical properties of a quantum dot (QD) can be used to control the spin state of individual \cite{Besombes2004,Kudelski2007,Kobak2014} or pairs \cite{Besombes2012,Krebs2013} of magnetic atoms. Including a magnetic atom in a QD offers additional degree of freedom. One can, for example, tune the environment of the localized spin by controlling the charge state of the QD. This electrical control can influence the magnetic anisotropy of the atom making these nano-sized systems attractive for basic investigations as well as for miniaturized data-storage applications.

In II-VI semiconductors, a Mn atom incorporated in a neutral self-assembled QD presents a small magnetic anisotropy of a few tens of $\mu eV$ resulting from a strained induced modification of the local crystal field \cite{Causa1980,Qazzaz1995}. This magnetic anisotropy is responsible for the Mn spin memory of a few $\mu$s observed at zero magnetic field \cite{LeGall2009,LeGall2010}. It has been shown recently that the fine structure splitting of the spin of a Mn can be modified optically \cite{LeGall2011,Jamet2013} or decreased significantly in strain-free QDs \cite{Besombes2014}. It is also expected that a Mn atom could develop a large anisotropy energy in the $meV$ range when it is exchange-coupled with a single confined heavy-hole spin \cite{Vyborny2012}. The two low energy hole-Mn states, with total spin $\Pi_z=\pm1$, should behave like an Ising spin system forming an atomic ferromagnet similar to what was observed for instance for single Co atoms deposited on a Cu surface \cite{Oberg2014}. In addition, in a semiconductor QD, the exchange induced Mn spin magnetic anisotropy could be electrically controlled by changing the Mn / heavy-hole overlap with an applied gate voltage.

In this paper, we analyze the dynamics of an individual Mn spin exchange coupled with a confined hole spin in CdTe/ZnTe self-assembled QDs. We first show that the complex formed by the Mn spin and the resident hole spin can be initialized within a few tens of nanoseconds under the optical injection of spin polarized carriers. We demonstrate that the spin pumping of the hole-Mn complex is controlled by the electron-Mn flip-flops induced by their exchange coupling within the positively charged exciton. This initialization process is modelled considering the exchange induced coherent dynamics of the electron-Mn complex in the excited state of the charged QD (positively charged exciton) and the dynamics of the hole-Mn complex in the ground state of the QD.

The transverse magnetic field dependence of the optical pumping signal reveals a magnetic anisotropy of the positively charged exciton-Mn complex. We show that this anisotropy is controlled by the strain induced fine structure of the Mn atom and a residual exchange coupling of the two holes with the Mn spin.

Using this optical pumping technique, we measure, at zero magnetic field and at T=7K, a spin relaxation time of the hole-Mn complex of about 1 $\mu s$. This is shorter than for a Mn spin in a neutral QD. We show that the source of this shortening of the spin life-time is the valence band mixing always present in strain induced II-VI self-assembled QDs.

The rest of this paper is organized as follows. In Sec. II we show how to extract all the parameters controlling the spin structure of positively charged Mn-doped QDs. In Sec. III we demonstrate that the optical injection of spin polarized carriers can be used to initialize the spin of the resident hole-Mn complex. In Sec. IV we describe the mechanism which is responsible for this efficient optical pumping. In Sec. V, we analyze the dynamics of the optical pumping and its excitation power and transverse magnetic field dependence. In Sec. VI, we present a model for the spin dynamics of positively charged QDs under optical injection of spin polarized carriers. Finally in Sec. VII, we present spin relaxation measurements and discuss the stability of the hole-Mn spin memory.

\section{Positively charged Mn-doped quantum dots.}

\begin{figure}[hbt]
\includegraphics[width=3.18in]{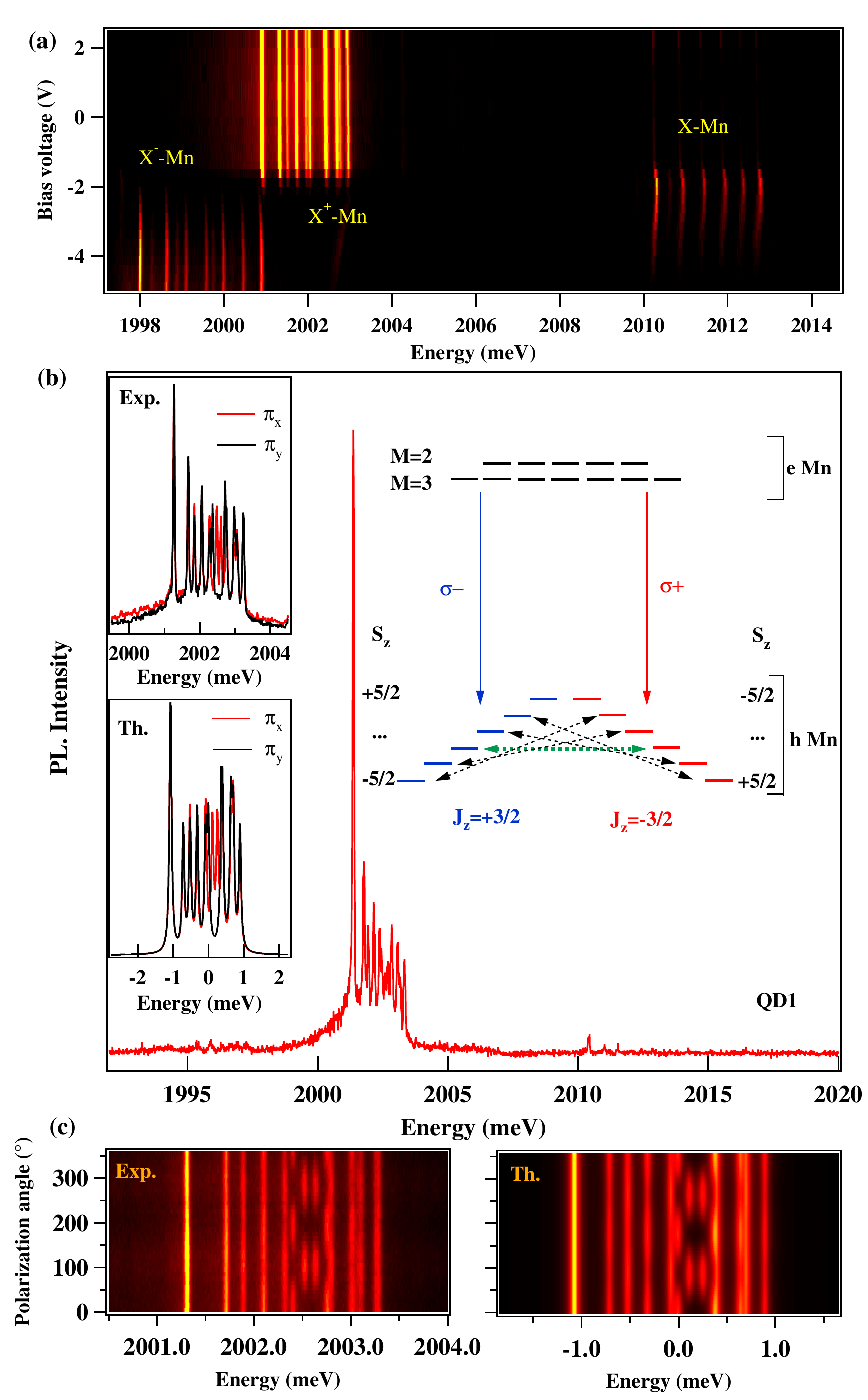}
\caption{(a) Color-scale plot of the PL intensity of a Mn-doped QD (QD1) in a Schottky structure as a function of emission energy and bias voltage for a CW excitation at $\lambda_{exc}=594nm$. The series of emission lines can be assigned to the recombination of the neutral exciton (X-Mn), positively charged exciton (X$^+$-Mn) and negatively charged exciton (X$^-$-Mn). (b) PL under co-circularly polarized excitation and detection of a positively charged Mn-doped QD. The left inset shows the experimental and calculated linearly polarized spectra along two orthogonal directions. The right inset is a simplified scheme of the energy levels of the ground (h-Mn) and excited states (e-Mn) of a positively charged Mn-doped QD. For h-Mn, the dotted arrows show the states coupled two by two by the valence band mixing. The states $|-1/2,\uparrow_e\rangle$ and $|+1/2,\downarrow_e\rangle$ (green arrow) are mixed and split. (c) Experimental (left) and calculated (right) color-scale plot of the dependence of the PL of X$^+$-Mn on the direction of a linear analyzer. Four lines in the center of the structure are linearly polarized. The parameters used in the model are listed in table \ref{table1}.}
\label{Fig1}
\end{figure}

Singly Mn-doped Cd(Mn)Te/ZnTe QDs containing a resident hole coming from a background p-type doping of the ZnTe barriers and ZnTe surface states can be isolated under quasi-resonant optical excitation below the ZnTe barrier energy. As presented in Fig.~\ref{Fig1}(a), the probability to find an individual confined holes in these QDs can be increased by applying a positive voltage between a back contact on the p-ZnTe substrate and a 6 nm thick semitransparent gold Schottky gate deposited on the surface of the sample.

Positively charged magnetic QDs can be identified by analysing the structure of their emission spectra. The emitting state in the X$^+$-Mn transition has two valence band holes and one electron coupled to the Mn. In a first approximation the effect of the two spin-paired holes on the Mn is zero. Thereby, the spin structure of the X$^+$-Mn state is governed by the ferromagnetic coupling between the electron and the Mn. The spin Hamiltonian of this system reads:
\begin{eqnarray}
{\cal H}_{eMn}=I_{eMn}\vec{S}\cdot\vec{\sigma}
\end{eqnarray}
\noindent with $I_{eMn}$ the exchange integral between the Mn and the electron with respective spins $\overrightarrow{S}$ and $\overrightarrow{\sigma}$. The twelve eigenstates of the electron-Mn complex are split into a ground state septuplet (total spin $M=3$) and a five-fold degenerate manifold with $M=2$ (see the inset of Fig.~\ref{Fig1}(b)). We label them all as $|M,M_z\rangle$.

The recombination of one of the hole with the electron of the $X^+$-Mn leaves a final state with a single hole coupled to the Mn. The hole-Mn exchange interaction is described by the spin Hamiltonian:
\begin{eqnarray}
{\cal H}_{hMn}=I_{hMn}\vec{S}\cdot\vec{J}
\end{eqnarray}
\noindent where I$_{hMn}$ is the exchange integral between the hole and the Mn atom. The hole spin operators, $\vec{J}$, represented in the basis of the two low energy heavy-hole states, are related to the Pauli matrices $\tau$ by $J_z= \frac{3}{2}\tau_z$ and $J_{\pm}= \xi \tau_{\pm}$ with $\xi=-2\sqrt{3}e^{-2i\theta}\rho_s/\Delta_{lh}$. $\rho_s$ is the coupling energy between heavy-holes and light-holes split by the energy $\Delta_{lh}$ and $\theta$ the angle relative to the [110] axis of the principal axis of the anisotropy responsible for the valence band mixing \cite{Fernandez2006,Leger2007}.

If we neglect the valence band mixing ($\rho_s/\Delta_{lh}$=0), the twelve eigenstates of ${\cal H}_{hMn}$ are organized as six doublets with well defined $S_z$ and $J_z$ (Mn and hole spin projection along the z axis). We label these states as $|S_z,J_z\rangle$. For each level of $X^+$-Mn with either $M=2$ or $M=3$, there are six possible final states after annihilation of an electron-hole pair. From this consideration alone, we would expect twelve spectrally resolved lines. As the Mn spin is not affected by the optical transition, the weight of each PL line is given by optical and spin conservation rules.

We consider, for instance, $\sigma+$ recombination where the $|\downarrow_e,\Uparrow_h\rangle$ e-h pair is annihilated. After the e-h annihilation, the resulting state is $|S_z,\Downarrow_h\rangle$ which is an eigenstate of ${\cal H}_{h-Mn}$. The intensity of the optical transition to a given final state $|S_z,\Downarrow_h\rangle$ is proportional to the overlap $\langle M,M_z|S_z,\downarrow_e\rangle$, which is nothing but a Clebsh-Gordan coefficient which appears in the composition of a spin 1/2 with a spin 5/2. The lowest energy transition, with $\sigma+$ polarization, would correspond to the high energy final state $|-5/2,\Downarrow_h\rangle$ and the initial state $|\Uparrow_h, \Downarrow_h\rangle\times |-5/2,\downarrow_e\rangle$ which is identical to $|3,-3\rangle$ and thereby gives the highest optical weight. In contrast, transition to that final state from $|2,M_z\rangle$ is forbidden. The other five doublets have optical weights lying between 1/6 and 5/6 with both $|2,M_z\rangle$ and $|3,M_z\rangle$ initial states. The number of spectrally resolved lines in this model is 11 and the PL of X$^+$-Mn can then be seen as a superposition of two substructures: six lines with intensities decreasing with increasing their energy position (transitions from  M=3 states) and five lines with intensities increasing with increasing their energy position (transitions from M=2 states).

As presented in Fig.~\ref{Fig1}(b) and \ref{Fig1}(c), the central lines of $X^+$-Mn are linearly polarized. This results from spin-flip interaction between the Mn and the hole induced by the presence of valence band mixing \cite{Leger2006}. Provided that $\rho_s/\Delta_{lh}<<1$, the effect of this interaction is small both on the wave function and on the degeneracy of all the hole-Mn doublets except the third (green arrow in the inset of Fig.~\ref{Fig1}(b)) which is split. The split states are the bonding and antibonding combinations of $|-1/2,\Uparrow_h\rangle$ and $|+1/2,\Downarrow_h\rangle$. These states are coupled, via linearly polarized photons, to the $|2,0\rangle$ and $|3,0\rangle$ electron-Mn states and four linearly polarized lines are observed on the emission spectra as shown in Fig.~\ref{Fig1}(c). Polarization directions are controlled by the QD anisotropy (strain and shape) responsible for the valence band mixing \cite{Leger2007}.

To describe in detail the emission spectra, we have also to take into account the perturbation of the wave function of the charged exciton in the initial state and of the hole in the final state by the hole-Mn exchange interaction. This perturbation depends on the value of the exchange energy between the Mn spin S$_z$ and the hole spin J$_z$ and it can be represented, using second order perturbation theory, by an effective spin Hamiltonian \cite{Besombes2005,Trojnar2013,Besombes2014}
\begin{eqnarray}
{\cal H}_{scat}=-\eta S_z^2
\end{eqnarray}
\noindent with $\eta>0$. The perturbation of the wave function is twice for the positively charged exciton where two holes interacts with the Mn. This perturbation is responsible for the irregular energy spacing of the X-Mn and X$_2$-Mn lines reported in previous work on self-assembled \cite{Besombes2005,Trojnar2013} and strain-free Mn-doped QDs \cite{Besombes2014}.

\begin{table}[htb] \centering
\caption{Values of the parameters used for the modelling of the five QDs discussed in this paper. I$_{eMn}$, I$_{hMn}$ and $\eta$ are expressed in $\mu eV$. For all the QDs, we use $\theta$=0. The width at half maximum of the depolarization curves obtained in the time resolved optical pumping measurements for X$^+$-Mn ($B^{X^+}_{1/2}$) and X-Mn ($B^{X}_{1/2}$) is expressed in Tesla. The relaxation time of the hole-Mn spin, $\tau_{r}$, is in $\mu s$.}
\label{table1}\renewcommand{\arraystretch}{1.0}
\begin{tabular}{p{1.0cm}p{1.0cm}p{1.0cm}p{1.0cm}p{1.0cm}p{1.0cm}p{1.0cm}}
\hline\hline
&  QD1 & QD2 & QD3 & QD4 & QD5 \\
\hline
I$_{eMn}$ & $-85$ & $-100$ & $-140$ & $-85$ & $-150$ \\
I$_{hMn}$ & $220$ & $230$ & $340$ & $195$ & $285$ \\
$\eta$  & $23$ & $17$ & $25$ & $25$ & $23$\\
$\rho_s/\Delta_{lh}$ & $0.06$ & $0.15$ & $0.07$ & $0.1$ & $0.17$\\
\hline
$B^{X^+}_{1/2}$ & $0.11$ & $0.07$ & $0.10$ & $0.13$ & $0.12$\\
$B^{X}_{1/2}$ & $0.07$ & $-$ & $0.06$ & $-$ & $-$\\
$\tau_{r}$ & $-$ & $1$ & $-$ & $0.9$ & $0.7$\\
\hline\hline
\end{tabular}
\end{table}

We can obtain numerical values of $I_{hMn}$, $I_{eMn}$, $\rho_s/\Delta_{lh}$ and $\eta$ comparing the transition probabilities calculated with the model to the experimental data (Fig.~\ref{Fig1}(c)). Values obtained for the five QDs discussed in this paper are listed in table \ref{table1}. A valence band mixing is observed in all the QDs but the value of the coefficient $\rho_s/\Delta_{lh}$ shows that the hole-Mn exchange interaction remains highly anisotropic. These values will be used in the model of the spin dynamics in the positively charged Mn-doped QDs.

\section{Evidence of the hole-Mn optical pumping.}

Similarly to what has been proposed \cite{Govorov2005} and recently demonstrated for an individual magnetic atom in neutral QDs \cite{LeGall2009,Goryca2009,LeGall2010,Baudin2011}, we show here that the optical injection of spin polarized carriers can be used to initialize the spin state of the hole-Mn complex.

\begin{figure}[hbt]
\includegraphics[width=3.5in]{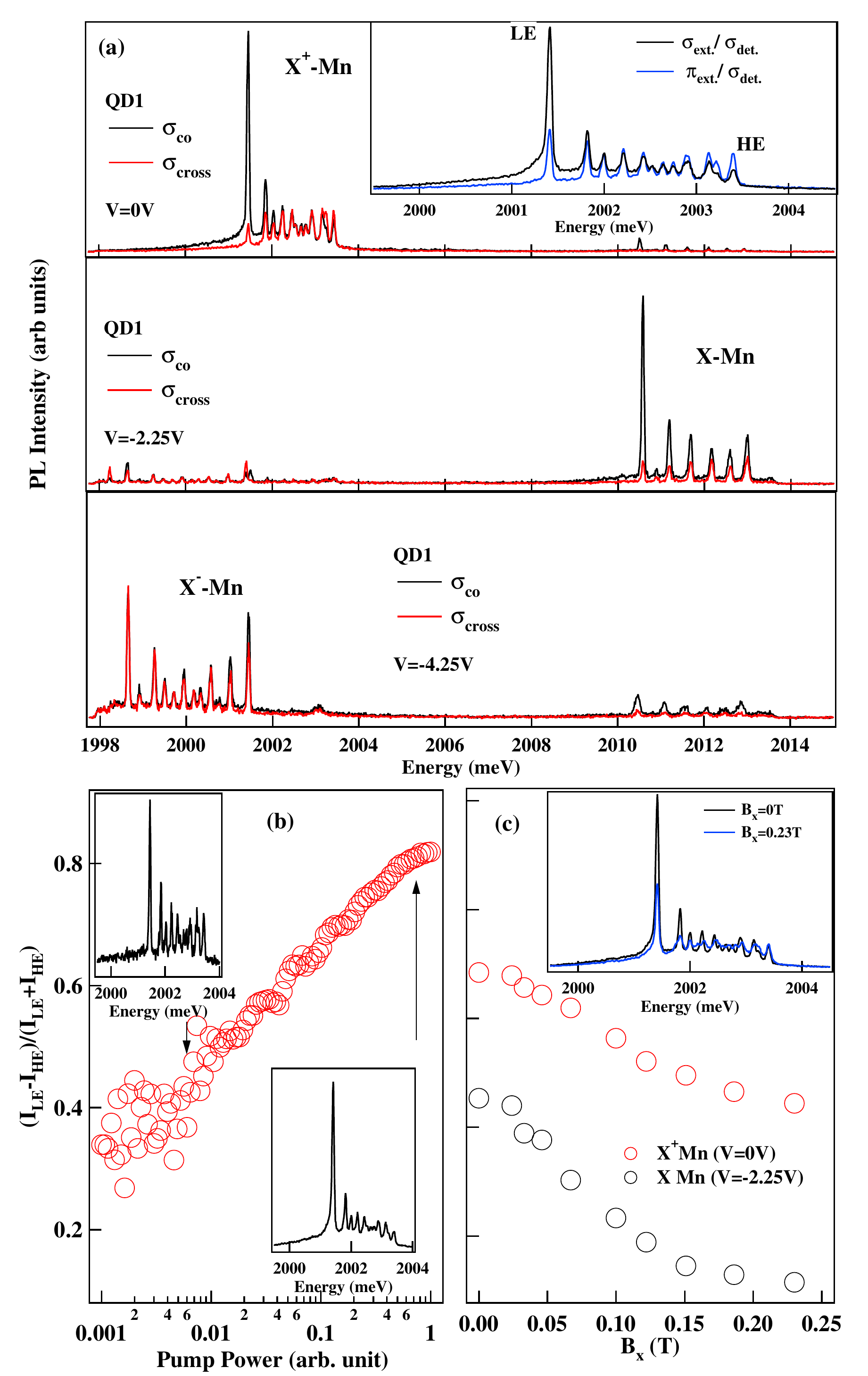}
\caption{(a) Co and cross-circularly polarized PL spectra of a Mn-doped QD (QD1) inserted in a Schottky structure for three different gate voltages. The top inset shows the circularly polarized PL spectra obtained under circular (black) and linear (blue) excitation. (b) Excitation power dependence of the ratio of intensity of the high and low energy lines $(I_{HE}-I_{LE})/(I_{HE}+I_{LE})$ of X$^+$-Mn under co-circularly polarized CW excitation and detection. (c) Transverse magnetic field dependence of the ratio of intensity of the high and low energy lines for X-Mn and X$^+$-Mn. The inset shows the co-circularly polarized spectra of X$^+$-Mn with and without transverse magnetic field.}
\label{Fig2}
\end{figure}

The emission spectra of a Mn-doped QD under circularly polarized continuous wave (CW) excitation below the ZnTe barrier \cite{Glazov2007} is presented in Fig.~\ref{Fig2}(a) for three different charge states of the dot ({\it i.e.} different gate voltages). For a positively charged or a neutral QD the emission is strongly co-circularly polarized with the excitation. In contrast, the emission of the negatively charged QD is almost unpolarized.

The polarization of the negatively charged QD is controlled by the spin of the injected hole. The weak circular polarization of X$^-$-Mn suggests that the hole spin is not conserved during the energy relaxation of the optically injected e-h pair or completely lost during the lifetime of the charged exciton. The strong co-circular polarization of X-Mn and X$^+$-Mn shows, on the other hand, that spin polarized electrons are efficiently injected in the QD and conserved during the lifetime of the exciton or the positively charged exciton.

As shown on the top inset of Fig.~\ref{Fig2}(a), the intensity distribution on the different lines on the emission of X$^+$-Mn strongly depends on the polarization of the excitation. For a linearly polarized excitation and a circular detection, the PL intensity reflects the oscillator strength of the transitions. For a circularly polarized excitation, most of the PL intensity is co-polarized and emitted on the low energy line of the X$^+$-Mn spectra. This highly out of equilibrium distribution of the PL intensity is a signature of the optical pumping of the hole-Mn complex.

The presence of optical pumping of the hole-Mn spin is confirmed by the excitation power dependence and the transverse magnetic field dependence of the intensity distribution in the positively charged exciton emission spectra. The normalized intensity ratio of the high energy (HE) and low energy (LE) lines of X$^+$-Mn under circularly polarized excitation (I$_{LE}$-I$_{HE}$)/(I$_{LE}$+I$_{HE}$) is presented in Fig.~\ref{Fig2}(b) as a function of the excitation power and in Fig.~\ref{Fig2}(c) as a function of a transverse magnetic field. The increase of the intensity ratio with the excitation power confirms the presence of a cumulative pumping process which is improved by the reduction of the delay between the successive injection of spin polarized carriers. On the other hand, the significant reduction of this intensity ratio in a weak transverse magnetic field is a signature of the precession of the spins involved in an optical pumping process.

With these optical pumping conditions, most of the light is emitted from the low energy line of X$^+$-Mn. The low energy PL line of X$^+$-Mn corresponds to a transition from the electron-Mn state $|3;+3\rangle=|+5/2,\uparrow_e\rangle$ to the hole-Mn state $|+5/2,\Uparrow_h\rangle$ in $\sigma-$ polarization and from the electron-Mn state $|3;-3\rangle=|-5/2,\downarrow_e\rangle$ to the hole-Mn state $|-5/2,\Downarrow_h\rangle$ in $\sigma+$ polarization. Consequently, the PL intensity distribution observed experimentally shows that just after the optical recombination, the hole-Mn complex is prepared in the spin state $|+5/2,\Uparrow_h\rangle$ under $\sigma-$ excitation and in the state $|-5/2,\Downarrow_h\rangle$ under $\sigma+$ excitation.

As illustrated in Fig.~\ref{Fig1}(b), the states $|+5/2,\Uparrow_h\rangle$ and $|-5/2,\Downarrow_h\rangle$ are the high energy states of the hole-Mn complex: an evolution of the system towards the ground states, $|-5/2,\Uparrow_h\rangle$ and  $|+5/2,\Downarrow_h\rangle$, will take place. A hole spin relaxation time of about 4 $ns$ was observed in CdTe/ZnTe QDs at zero magnetic field using as a probe the time evolution of the polarization rate of the negatively charged exciton \cite{LeGall2012}. The hole spin relaxation can even be faster in the presence of an external applied magnetic field: the interaction with acoustic-phonons is enhanced when the spin states of the hole are split by a few hundreds $\mu eV$  \cite{Woods2004,Roszak2007,Cao2011,Bounouar2012}. Similarly, the hole spin split by the exchange interaction with the Mn is expected to relax towards the low energy states in a $ns$ time scale. On the other hand, a Mn spin relaxation of a few $\mu s$ is observed for a Mn spin at zero magnetic field \cite{LeGall2009} and this relaxation time is even longer under magnetic field \cite{Goryca2009}. We can then expect that the Mn spin is conserved during the hole spin relaxation. After this relaxation, the hole-Mn spin is prepared in one of the two low energy states, either $|+5/2,\Downarrow_h\rangle$ for a $\sigma-$ excitation or $|-5/2,\Uparrow_h\rangle$ for a $\sigma+$ excitation.

\section{Mechanism of the hole-Mn optical pumping.}

\begin{figure}[hbt]
\includegraphics[width=3.25in]{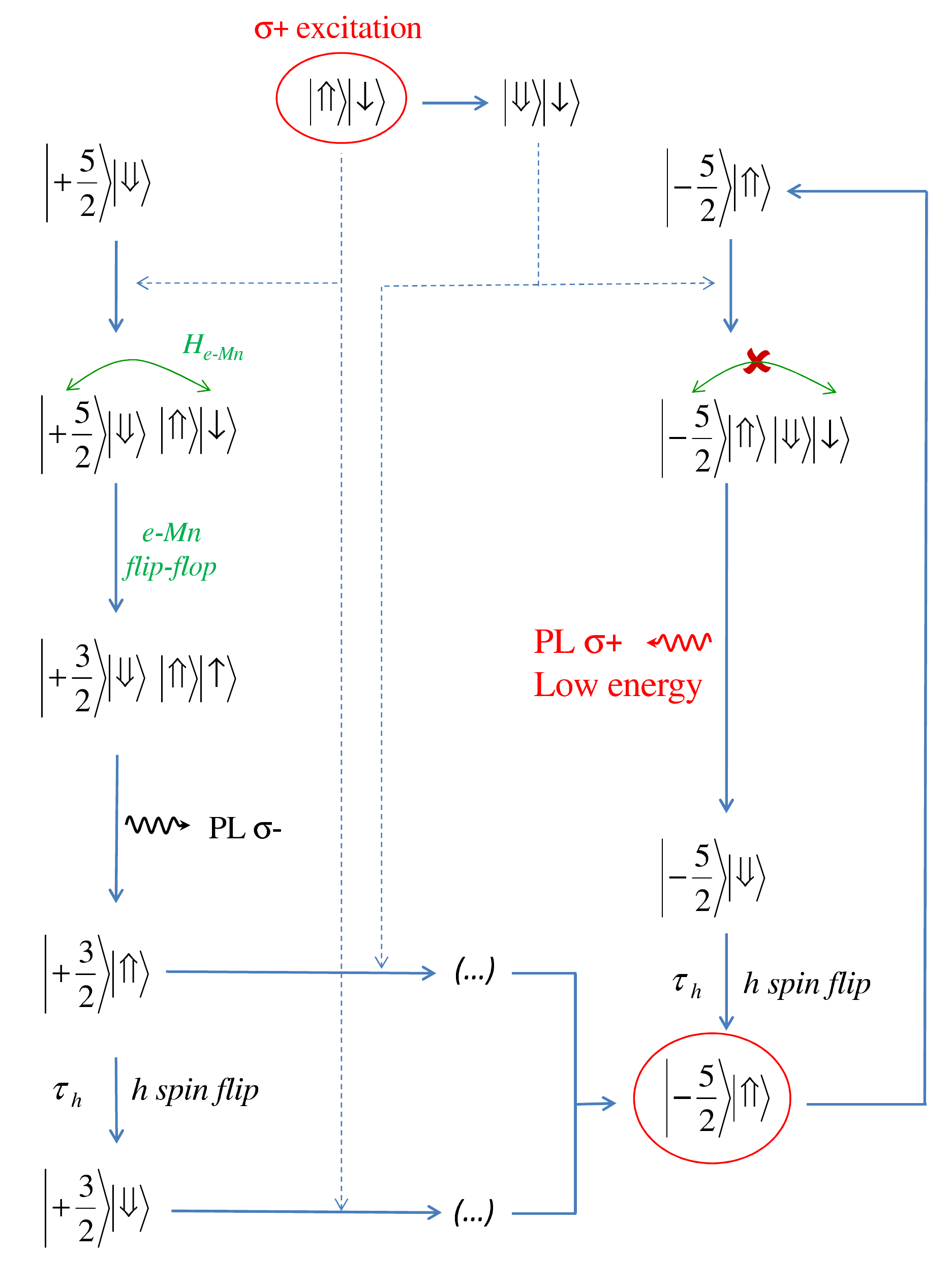}
\caption{Scheme of the optical pumping mechanism of the spin of the hole-Mn complex under $\sigma$+ excitation. The electron-Mn flip-flops controlled by ${\cal H}_{e-Mn}$ and the electron-Mn coherence time $T_{2}^{eMn}$ prepare the Mn spin in the state $S_z=-5/2$. See the text for a detailed description of the pumping process.}
\label{Fig3}
\end{figure}

A scheme of the optical pumping process that we propose for the initialization of the hole-Mn system is presented in Fig.~\ref{Fig3} for a $\sigma+$ excitation. We consider first that the hole-Mn complex is initially in one of the two ground states, either $|+5/2,\Downarrow_h\rangle$ or $|-5/2,\Uparrow_h\rangle$, after thermalisation in the dark. This hypothesis is made for clarity of the demonstration: starting from any other hole-Mn spin state would lead to the same conclusion.
It is assumed that, after the injection of excitons by the optical excitation, the spin of the hole can flip but the spin of the electron is conserved. This is consistent with the polarisation rates observed in Fig.~\ref{Fig2}(a). Consequently, for a $\sigma+$ optical excitation, either bright excitons $|+1\rangle=|\Uparrow_h,\downarrow_e\rangle$ or dark ones $|-2\rangle=|\Downarrow_h,\downarrow_e\rangle$ can be captured by the QD.

Let's consider first the situation where the hole-Mn complex is initially in the state $|+5/2,\Downarrow_h\rangle$. In this situation, only a bright exciton can be captured by the dot to form the positively charged exciton $|+5/2,\Downarrow_h\rangle\times|\Uparrow_h,\downarrow_e\rangle$. The isotropic coupling between the electron and the Mn induces electron-Mn spin flip-flops. After one flip-flop, the spin up electron can recombine with the resident spin down hole emitting a $\sigma-$ photon. The hole-Mn complex is left in the state $|+3/2,\Uparrow_h\rangle$. This high energy hole-Mn state can relax within a few hundreds $ps$ to the lower energy state $|+3/2,\Downarrow_h\rangle$ though a hole spin-flip induced by acoustic phonons. Then, a second bright exciton can be injected in the dot containing a spin down hole. An electron-Mn flip-flop followed by the recombination of a $\sigma-$ exciton leave the hole-Mn complex in the state $|+1/2,\Uparrow_h\rangle$. Alternatively, a dark exiton can be captured before the hole spin relaxation. An electron-Mn flip-flop and an exciton recombination in $\sigma-$ polarization will also leave the hole-Mn spin in the state $|+1/2,\Uparrow_h\rangle$. Repeating five times this process of injection of spin polarized electron (either $|+1\rangle$ or $|-2\rangle$ exciton) with electron-Mn flip-flops will finally drive the hole-Mn system from the initial state $|+5/2,\Downarrow_h\rangle$ to the state $|-5/2,\Uparrow_h\rangle$.

When the hole-Mn complex is initially in the state $|-5/2,\Uparrow_h\rangle$, the QD can only capture a dark exciton. The positively charged exciton is in the state $|-5/2,\Uparrow_h\rangle\times|\Downarrow_h,\downarrow_e\rangle$. The electron-Mn flip-flops are blocked because both the electron and Mn spin $z$ components are minimum. The injected spin down electron can recombine with the resident spin up hole emitting a co-circularly polarized $\sigma+$ photon on the low energy line of X$^+$-Mn. After recombination, the hole-Mn system is in the state $|-5/2,\Downarrow_h\rangle$. After a fast hole spin relaxation, the hole-Mn complex returns to the state $|-5/2,\Uparrow_h\rangle$. Under CW $\sigma+$ excitation, this process of capture of a dark exciton followed by an optical recombination and a spin flip of the hole is repeated, and the hole-Mn complex remains blocked in the state $|-5/2,\Uparrow_h\rangle$.

The sequence presented here for a $\sigma+$ excitation leads to the preparation of the hole-Mn complex in the state $|-5/2,\Uparrow_h\rangle$. Therefore, as observed in the experiment, under $\sigma+$ excitation the positively charged QD emits predominantly co-circularly polarized $\sigma+$ photons on the low energy line of X$^+$-Mn.

A similar scheme can be obtained under $\sigma-$ excitation. In this configuration, after the pumping cycle induced by the injection of bright and dark excitons with $|\uparrow_e\rangle$ electrons, the hole-Mn spin is blocked in the state $|+5/2,\Downarrow_h\rangle$ and the QD emits co-circularly polarized $\sigma-$ photons on the low energy line of X$^+$-Mn.

The proposed optical pumping scheme is compatible with all the experimental observations under CW circularly polarized excitation. It suggests that the optical pumping of the hole-Mn spin is induced by the exchange interaction of the Mn spin with the injected spin polarized electrons. Its dynamics should be controlled by ${\cal H}_{eMn}$ and the generation rate of excitons.

\section{Dynamics of the hole-Mn optical pumping.}

To extract the dynamics of this optical pumping process, we modulate the circular polarisation of the CW excitation with and electro-optic modulator combined with a quarter-wave plate (circular polarization switching time of about 10 ns) and analyze the time evolution of the intensity distribution on the X$^+$-Mn spectra.

\begin{figure}[hbt]
\includegraphics[width=3.5in]{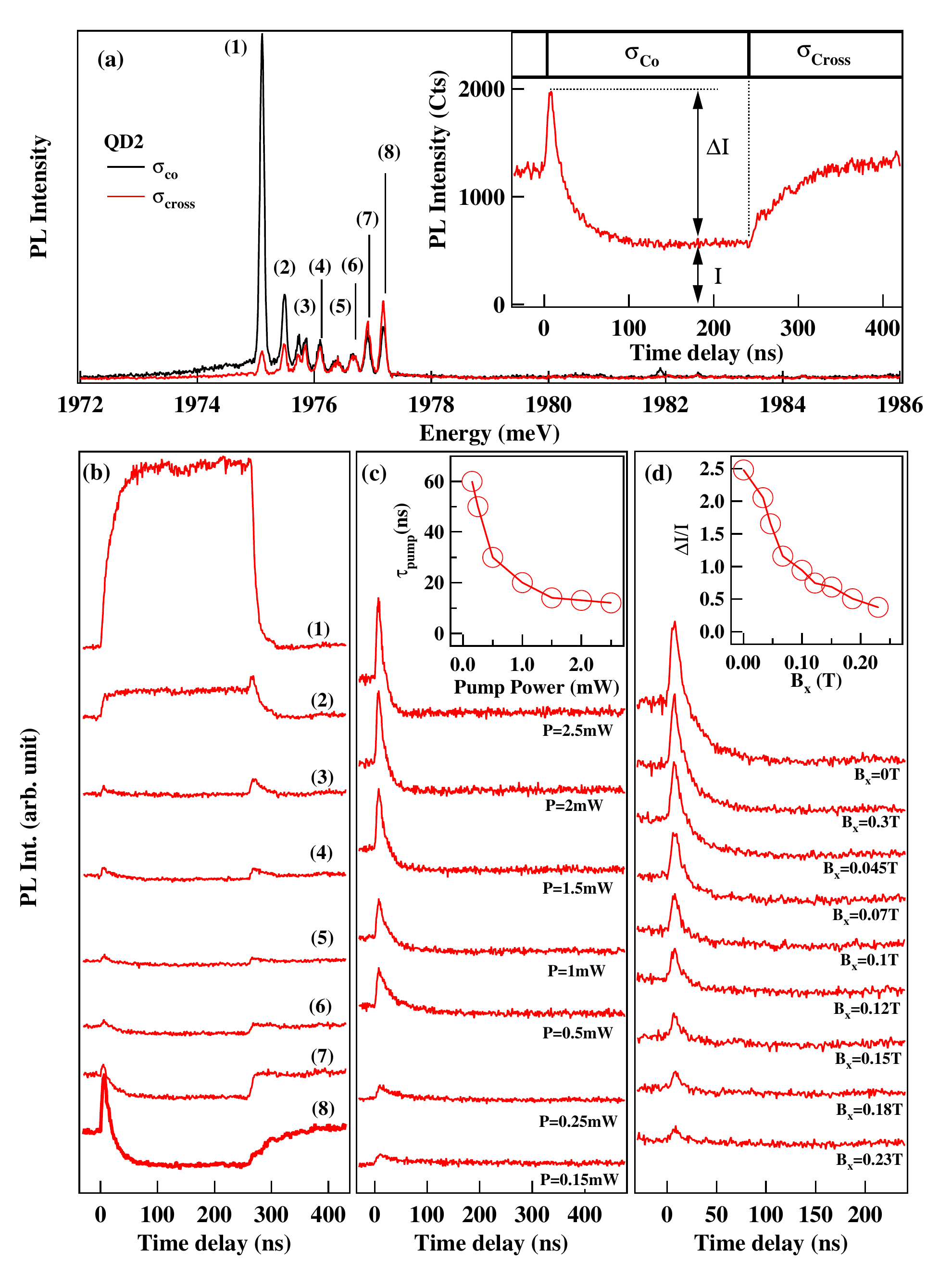}
\caption{(a) Co and cross-circularly polarized PL of X$^+$-Mn in QD2. The inset shows the time evolution of the PL of the hight energy line (8) under polarization switching of the excitation. (b) Time evolution of the circularly polarized PL of the different lines of X$^+$-Mn (labelled from (1) to (8)) under polarization switching of the excitation. (c) Excitation power dependence of the PL transient measured on the high energy line (8). The inset shows the power dependence of the pumping time. (d) Transverse magnetic field dependence of the PL transient measured on the high energy line of X$^+$-Mn. The inset shows the magnetic field dependence of the amplitude of the optical pumping transient $\Delta I/I$.}
\label{Fig4}
\end{figure}

The time evolution of the circularly polarized PL of each lines of a X$^+$-Mn that can be spectrally resolved are presented in Fig.~\ref{Fig4}(b). In the co-circular polarization configuration for the excitation and the detection we observe two PL intensity transients: first an abrupt one limited by the 10 ns time resolution of the polarization switching. It reflects the population change of the spin-polarized exciton. Then, a slower transient is observed with opposite signs on the two extreme PL lines (lines 1 and 8 which correspond to final states with parallel or anti-parallel hole and Mn spins). The characteristic time of this exponetial transient in the tens of $ns$ range is, at low excitation power, inversely proportional to the pump intensity and then saturates (Fig.~\ref{Fig4}(c)). This is a signature of the optical pumping process which realizes an orientation of the hole-Mn spin.

As presented in Fig.~\ref{Fig4}(d), an in-plane magnetic field, B$_x$, of a few tens of milli-Tesla induces a decrease of the efficiency of the optical pumping as measured on the amplitude of the pumping PL transient of the high energy line (see inset of Fig.~\ref{Fig4}(a)). This results is at first sight surprising as one could expect a large magnetic anisotropy for the hole-Mn system controlled by the large anisotropy of the hole $g$ factor.

\begin{figure}[hbt]
\includegraphics[width=3.25in]{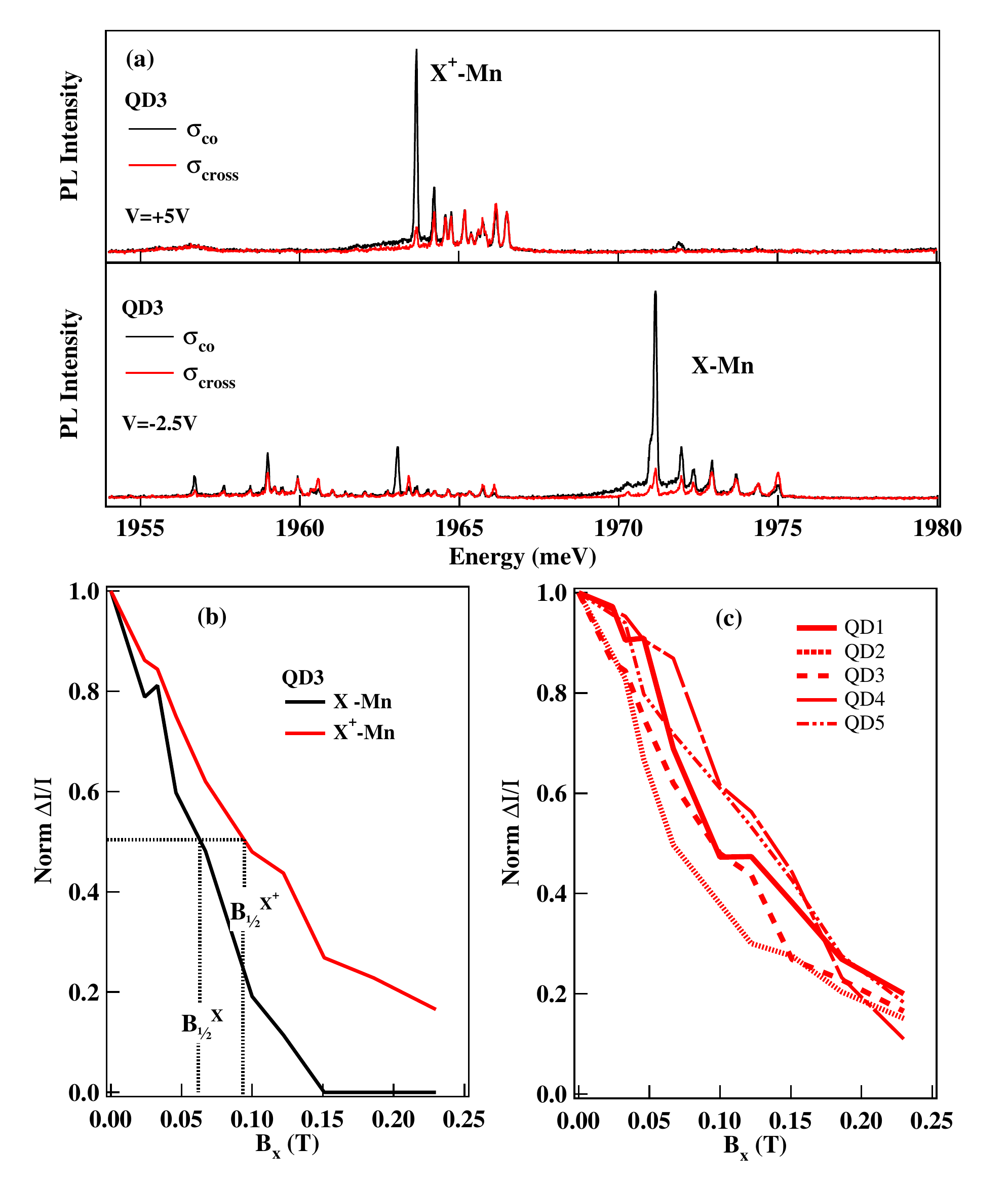}
\caption{(a) Co and cross-circularly polarized PL of X$^+$-Mn and X-Mn in a Mn-doped QD (QD3). (b) Transverse magnetic field dependence of the amplitude of the optical pumping transients measured on the high energy line of X-Mn and on the high energy line of X$^+$-Mn in QD3. (c) Transverse magnetic field dependence of the amplitude of the optical pumping transients measured on the high energy line of X$^+$-Mn compared for five different QDs from QD1 to QD5. The corresponding half width at half maximum of the depolarization curves are listed in table \ref{table1}.}
\label{Fig5}
\end{figure}

Under CW excitation, the QD is occupied a significant part of the time by a positively charged exciton. Within the X$^+$-Mn complex, the coupling of the Mn spin with the carriers is dominated by the exchange interaction with the electron which results in an isotropic hybrid spin system. In the absence of magnetic anisotropy of the Mn spin, this hybrid spin could freely precess in a transverse weak magnetic field suppressing any possibility of spin transfer from the injected spin polarized electron to the hole-Mn system. However, it is known from previous studies of spin dynamics in Mn-doped neutral QDs that the ground state of the Mn presents a fine structure. In epitaxial structures, this fine structure is dominated by a magnetic anisotropy with an easy axis along the QD axis due to built-in strains in the QD plane. The fine structure splitting of the Mn is described by the spin Hamiltonian:

\begin{eqnarray}
{\cal H}_{Mn}=D_0S^2_z+E[S_x^2-S_y^2]+\frac{1}{6}a[S_x^4+S_y^4+S_z^4]
\end{eqnarray}

\noindent where $D_0$ is proportional to the biaxial strain, $E$ describes an anisotropy of the strain in the QD plane and $a$ is the intrinsic splitting induced by the tetrahedral environment of the Mn in the crystal ($a=0.32\mu eV$ for a Mn in CdTe). The magnetic anisotropy term $D_0$ is responsible for the transverse magnetic field dependence of the optical pumping of the Mn observed in a neutral QD \cite{LeGall2009}. $D_0$ can vary from 0 $\mu$eV for a strain free QD \cite{Besombes2014} to about 12 $\mu$eV for a fully strained CdTe layer matched on a ZnTe substrate \cite{LeGall2009}. This magnetic anisotropy term influences the electron-Mn spin dynamics in a weak transverse magnetic field.

Within the positively charged exciton, the weak interaction of the Mn with the two holes, described by ${\cal H}_{scat}$, also introduces a magnetic anisotropy. The influence of this parameter is revealed by the comparison of the transverse magnetic field dependence of the optical pumping efficiency in neutral and positively charged QDs. The transverse magnetic field dependence of the pumping transient detected on the high energy line of X-Mn and on the high energy line of X$^+$-Mn for the same QD are presented in Fig.~\ref{Fig5}(b). The half width at half maximum of the depolarization curve for X$^+$-Mn ($B^{X^+}_{1/2}$) is significantly larger than for X-Mn ($B^{X}_{1/2}$). The values of $B^{X^+}_{1/2}$ and $B^{X}_{1/2}$ measured for the different QDs studied in thispaper are listed in Table \ref{table1}. Whereas the depolarization curve of X-Mn is only controlled by $D_0$ \cite{LeGall2009}, the total anisotropy of the electron-Mn system induced by both $D_0$ and $\eta$ is responsible for the transverse magnetic field dependence of the optical pumping signal of X$^+$-Mn.

\section{Model of the hole-Mn optical pumping.}

We propose a model to better understand this optical pumping mechanism and its transverse magnetic field dependence. It is based on the calculation of the time evolution of the populations of the twelve X$^+$-Mn states in the excited state of the QD and the twelve hole-Mn states in the ground state. In this model, we neglect the hyperfine coupling between the electronic and nuclear spin (I=5/2) of the Mn atom which is, in the investigated QDs, weaker than the electron-Mn and hole-Mn exchange interactions. We use the density matrix formalism (24 x 24 density matrix $\rho$) to calculate the population in this multi-level system. The master equation which governs the evolution of $\rho$ can be written in the Lindblad form as:

\begin{equation}
{\frac{\partial \rho}{\partial t}=-i/\hbar[{\cal H},\rho]+L\rho}
\end{equation}

\noindent where ${\cal H}$ is the Hamiltonian of the complete system ($X^+$-Mn and hole-Mn) and $L\rho$ describes the coupling or decay channels resulting from an
interaction with the environment \cite{Exter2009,Roy2011}. The coherent evolution of this multi-level system is controlled by the Hamiltonian ${\cal H}_{Mn}+{\cal H}_{eMn}+2{\cal H}_{scat}$ for $X^+-Mn$ and ${\cal H}_{Mn}+{\cal H}_{hMn}+{\cal H}_{scat}$ for the hole-Mn complex. For the magnetic field dependence, Zeeman terms for the hole, the electron and the Mn are also included.

\begin{figure}[hbt]
\includegraphics[width=3.5in]{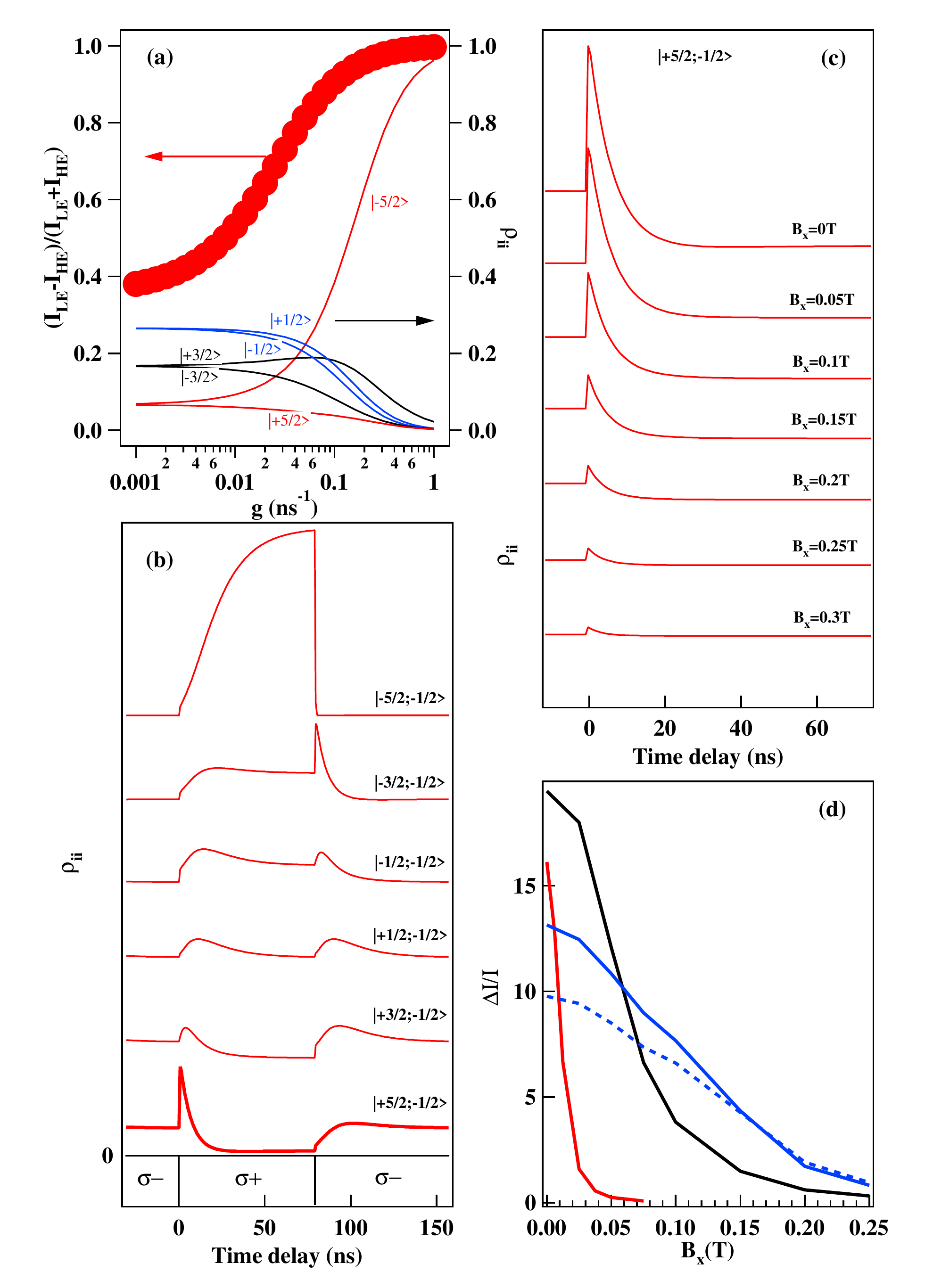}
\caption{Optical pumping signal calculated with the parameters of QD2 (see table \ref{table1}). (a) Calculated intensity ratio of the high and low energy lines of X$^+$-Mn and population of the Mn spin states presented as a function of the generation rate of excitons under $\sigma+$ excitation ($g=g_{|+1\rangle}=g_{|-2\rangle}=1/\tau_{g}$). (b) Time evolution of the population of the different electron-Mn states under excitation with alternate circular polarization ($\tau_{g}=4ns$). The curves are shifted for clarity. (c) Transverse magnetic field dependence of the pumping transient observed on the high energy line ($\tau_{g}=4ns$). For (a), (b) and (c) the parameters used in the calculation are $\tau_{Mn}=5\mu s$, $\tau_{h}=1ns$, $\tau_r=0.3ns$, $T_{2}^{hMn}=0.5 ns$, $T_{2}^{eMn}=0.1 ns$, T=10K, $a=0.32 \mu eV$, D$_0$=6$\mu eV$, E=0.5$\mu eV$, $g_{Mn}=2$, $g_{h}=0.6$ and $g_{e}=-0.4$. (d) Amplitude of the optical pumping transient as a function of transverse magnetic field calculated with the previous parameters (black), with D$_0$=0$\mu eV$, E=0$\mu eV$, $\eta=0\mu eV$ (red), with $T_{2}^{eMn}=\infty$ (i.e. lifetime limited coherence of e-Mn) (blue) and with $T_{2}^{eMn}=\infty$, $T_{2}^{hMn}=1 ns$ (dotted blue).}
\label{Fig6}
\end{figure}

In agreement with the polarized PL measurements presented in Fig.\ref{Fig2}(a), we consider that we generate spin polarized electrons and random holes ({\it i.e.} a $\sigma+$ excitation generates $|+1\rangle$ excitons or $|-2\rangle$ excitons with the same probability). The only recombination channel for the optically created $X^+-Mn$ is a radiative recombination with a characteristic lifetime $\tau_r=0.3ns$ which conserves the Mn spin.

We take into account a spin relaxation time of the Mn in the exchange field of the hole, $\tau_{Mn}$, describing relaxation channels with a change of the Mn spin by one unit. We choose $\tau_{Mn}=5\mu s$ which is the typical value of relaxation time for a Mn spin in an empty QD at zero magnetic field and we neglect here the possible decrease of this relaxation time under the injection of free carrier in the vicinity of the QD \cite{Besombes2008}. We include a relaxation of the hole spin coupled with the Mn, $\tau_h$, in the $ns$ range. The transition rates $\Gamma_{\gamma\rightarrow\gamma'}$ between the different states of the hole-Mn complex depend on their energy separation $E_{\gamma\gamma'}=E_{\gamma'}-E_{\gamma}$. Here we use $\Gamma_{\gamma\rightarrow\gamma'}$=1/$\tau_{i}$ if $E_{\gamma\gamma'}<0$ and $\Gamma_{\gamma\rightarrow\gamma'}$=1/$\tau_{i}e^{-E_{\gamma\gamma'}/k_BT}$ if $E_{\gamma\gamma'}>0$ ($\tau_i$ corresponding either to $\tau_{Mn}$ or $\tau_{h}$) \cite{Govorov2005}. This describes a thermalization among the twelve hole-Mn levels. A pure dephasing time for the hole-Mn system ($T_2^{hMn}$) and for the electron-Mn system ($T_2^{eMn}$) are also included in the model. The incoherent optical excitation, the optical recombination, the relaxation and pure dephasing terms are inserted in $L\rho$ following the method presented in reference \cite{Jamet2013}.

With this level scheme, we can calculate the population of each X$^+$-Mn state ({\it i.e} electron-Mn states $|S_z,\sigma_z\rangle$) under CW or alternate circularly polarized excitation. These populations are proportional to the intensity of the X$^+$-Mn PL lines. For instance, in $\sigma+$ polarization, the low energy line corresponds to the transition from the low energy electron-Mn state $|3;-3\rangle=\frac{1}{\sqrt{6}}[\sqrt{6}|-5/2,\downarrow_e\rangle$] to the high energy hole-Mn state $|-5/2,\Downarrow_h\rangle$ and the high energy line corresponds to the transition from the high energy level $|2;+2\rangle=\frac{1}{\sqrt{6}}[\sqrt{1}|+3/2,\uparrow_e\rangle-\sqrt{5}|+5/2,\downarrow_e\rangle$] to the low energy level $|+5/2,\Downarrow_h\rangle$. The normalized intensity ratio of the low and high energy lines, which measures the difference of population of the Mn spin states $S_z=-5/2$ and $S_z=+5/2$, is then given by:
\begin{eqnarray}
\frac{I_{LE}-I_{HE}}{I_{LE}+I_{HE}}=\frac{\rho_{|-5/2,\downarrow_e\rangle}-5/6\rho_{|+5/2,\downarrow_e\rangle}}{\rho_{|-5/2,\downarrow_e\rangle}+5/6\rho_{|+5/2,\downarrow_e\rangle}}
\end{eqnarray}

The calculated excitation power dependence of the intensity ratio of the high and low energy lines is presented in Fig.~\ref{Fig6}(a) for the exchange parameters of QD2 (see table \ref{table1}) and a CW $\sigma+$ excitation. As observed in the experiment (Fig.~\ref{Fig2}(b)), a variation of three orders of magnitude of the exciton generation rate is necessary to probe all the range of variation of the PL intensity ratio. At high excitation intensity, the population of the different Mn spin states $|S_z\rangle$ deduced from the model confirms that the Mn spin is highly polarized with, for a $\sigma+$ excitation, most of the population in the spin state $S_z=-5/2$ (Fig.~\ref{Fig6}(a)).

Fig.~\ref{Fig6}(b) presents the calculated time evolution of the population of the electron-Mn states giving rise to a $\sigma+$ emission (spin states $|S_z,\downarrow_e\rangle$) under CW excitation with alternate circular polarization. The intensity of the low and high energy lines of X$^+$-Mn are given by $\rho_{|-5/2,\uparrow_e\rangle}$ and $5/6\rho_{|+5/2,\downarrow_e\rangle}$ respectively. The main feature of the time resolved optical pumping experiments (see Fig.~\ref{Fig4}(b)) are well reproduced by the model. For instance, under $\sigma +$ excitation and $\sigma+$ detection, a large increase of the intensity of the low energy line is obtained together with a decrease of the intensity of the high energy line. This corresponds to the progressive transfer of population from $S_z=+5/2$ to $S_z=-5/2$ by the pumping process induced by the optical injection of $|\downarrow_e\rangle$ electrons and electron-Mn flip-flops controlled by ${\cal H}_{eMn}$ and $T_{2}^{eMn}$.

The influence of a transverse magnetic field, $B_x$, on the optical pumping transient can also be qualitatively described by this model. A transverse magnetic field dependence of the optical pumping transient calculated for the high energy line of X$^+$-Mn is presented in Fig.~\ref{Fig6}(c). As in the experiment, a significant decrease of the amplitude of the pumping transient is observed for a transverse field of a few tens of mT.

Calculated depolarization curves ({\it i.e} transverse magnetic field dependence of the parameter $\Delta I/I$ defined in Fig.~\ref{Fig4}(a)) are presented in Fig.~\ref{Fig6}(d) for different values of the parameters $D_0$, $\eta$, $T_{2}^{hMn}$ and $T_{2}^{eMn}$. In the absence of magnetic anisotropy ($D_0$=0 and $\eta$=0, red line in Fig.~\ref{Fig6}(d)), the transverse magnetic field dependence is determined by the precession of the electron and Mn spin interrupted by a dephasing process (controlled by $T_{2}^{eMn}$): a transverse field of 20 $mT$ completely destroys the optical pumping. In the presence of $D_0$ and $\eta$ (black line in Fig.~\ref{Fig6}(d)) the precession is blocked in a weak transverse magnetic field and the optical pumping can take place. A transverse field of 200 $mT$ is now required to completely suppress the optical pumping. $D_0$ and $\eta$, that both introduce an anisotropy in the electron-Mn system, appear as the main parameters that control the depolarization curve of X$^+$-Mn. The influence of these two independent parameters that changes from dot to dot is responsible for the dispersion of $B^{X^+}_{1/2}$ measured in different dots (Fig.~\ref{Fig5}(c)). The influence of $\eta$ on the magnetic anisotropy of $X^+-Mn$ also explains the difference of magnetic anisotropy measured for $X^+-Mn$ and $X-Mn$, the latter being only influenced by $D_0$ (Fig.~\ref{Fig5}(b)).

This modelling also shows that the amplitude of the pumping transient and its transverse magnetic field dependence are influenced by the value of the coherence time of the electron-Mn and hole-Mn systems (Fig.~\ref{Fig6}(d)). These dephasing times are not directly accessible from our measurements. Despite the determination of $\eta$ from a modelling of the PL of $X^+-Mn$, we cannot independently determine $D_0$ from the depolarization curves. Nevertheless, the shape of the depolarization curve of QD2 (Black curve in Fig.~\ref{Fig4}(d)) can be well reproduced by the model with reasonable order of magnitudes for all these parameters: $\eta=17\mu eV$ (obtained from a modelling of the PL of $X^+-Mn$ in QD2), $T_{2}^{hMn}=0.5 ns$, $T_{2}^{eMn}=0.1 ns$ and $D_0=6\mu eV$ (partial relaxation of the strain at the Mn location).

\section{Hole-Mn spin relaxation.}

\begin{figure}[hbt]
\includegraphics[width=3.25in]{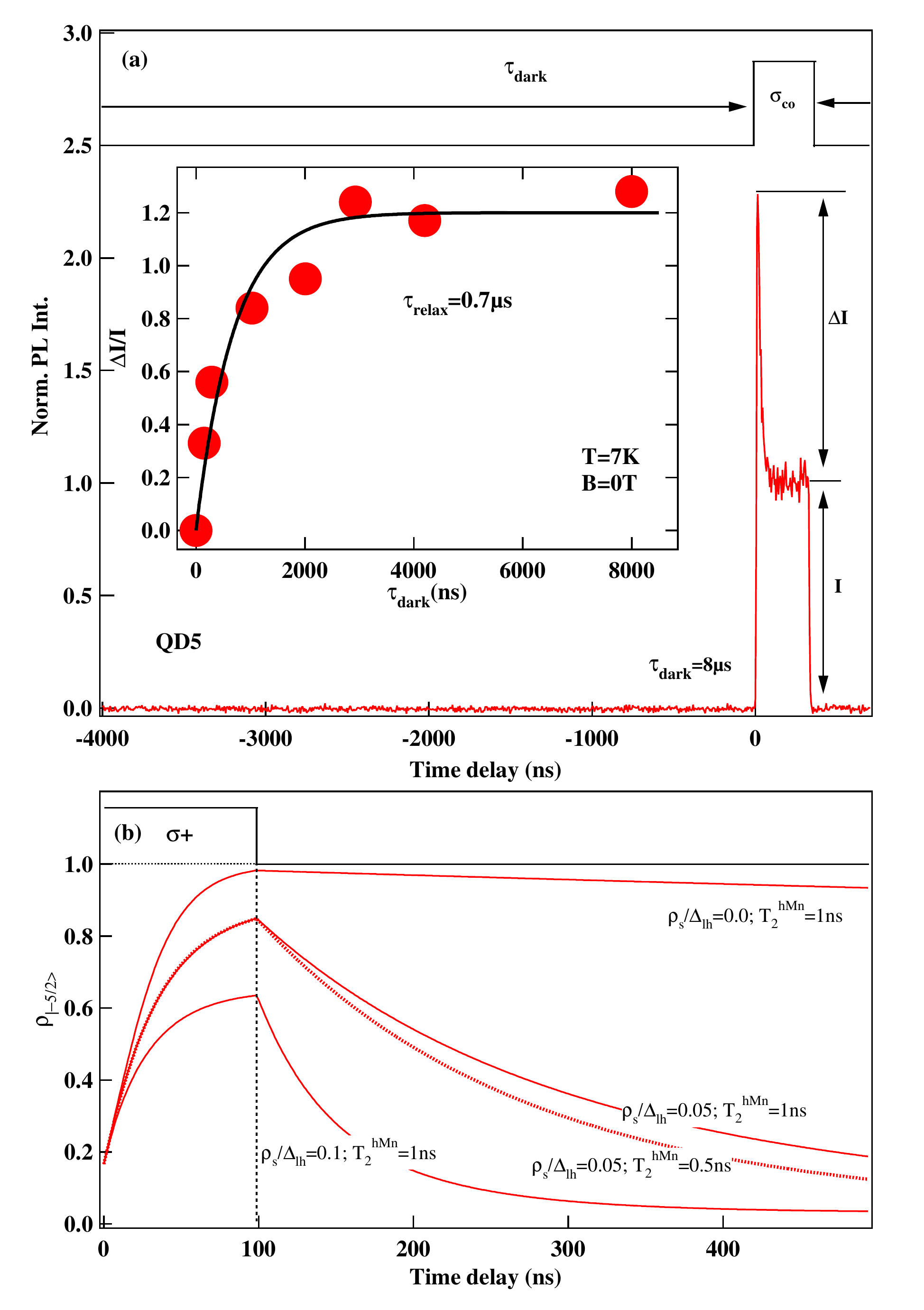}
\caption{(a) Optical pumping transient obtained on the high energy line of X$^+$-Mn in QD5 under excitation with circularly polarized light trains of 300 ns separated by a dark time of $\tau_{dark}$=8$\mu$s (the excitation sequence is displayed on the top). The inset shows the evolution of the amplitude of the optical pumping transient with $\tau_{dark}$ for B = 0T and T=7K. (b) Calculated time evolution in the dark of the population of the Mn spin state $S_z=-5/2$ initially pumped by a $\sigma+$ CW excitation (excitation sequence displayed at the top). The calculation are performed with the exchange couplings parameters of QD5 (see table \ref{table1}), $\tau_{Mn}=5\mu s$, $\tau_{h}=1ns$, $\tau_r=0.3ns$, $T_{2}^{eMn}=0.1 ns$, T=10K, $a=0.32 \mu eV$, D$_0$=6$\mu eV$, E=0.5$\mu eV$ and different amplitude of valence band mixing $\rho_s/\Delta_{lh}$ and hole-Mn coherence time $T_{2}^{hMn}$.}
\label{Fig7}
\end{figure}

Having established a method to initialize the spin of the hole-Mn complex and identified the mechanism responsible for the optical pumping, we performed pump-probe experiments to observe how the hole-Mn spin is conserved in the absence of optical excitation. In this experiment (Fig.~\ref{Fig7}(a)), the QD is excited with trains of 300 ns circularly polarized pulses separated by a variable dark time ($\tau_{dark}$). At the end of the excitation pulse, an out of equilibrium population for the hole-Mn system is prepared by optical pumping.

We measure after the time $\tau_{dark}$ the amplitude of the optical pumping transient observed at the beginning of the excitation pulse. The presence of this pumping transient reflects a partial relaxation of the hole-Mn polarization during the dark time. The amplitude of the transient saturates as $\tau_{dark}$ increases, demonstrating the full relaxation of the hole-Mn spin. An exponential fit of the variation of this amplitude with $\tau_{dark}$ is used to estimate the relaxation time of the hole-Mn spin at zero magnetic field and T=7K. A relaxation time of about 0.7 $\mu s$ is obtained for the example presented in Fig.~\ref{Fig7}(a) (QD5). We always found a relaxation time in the 1 $\mu s$ range or shorter for all the studied positively charged QDs (see table \ref{table1}). This relaxation time for the Mn coupled with a confined hole spin is shorter than the typical relaxation time observed at zero magnetic field for a Mn atom in a neutral QD (about 5$\mu s$) \cite{LeGall2009}.

The exchange interaction of the Mn with a heavy-hole spin creates a large magnetic anisotropy for the Mn and one could expect to increase the Mn spin memory at zero magnetic field. However, in the presence of valence band mixing, ${\cal H}_{h-Mn}$ couples two by two the different hole-Mn levels (inset of  Fig.~\ref{Fig1}(b)). This coupling induces a transfer of population between the different hole-Mn levels. The transfer of population becomes irreversible in the presence of dephasing in the hole-Mn system and could significantly contribute to the hole-Mn spin relaxation.

To analyze more in detail this possible spin relaxation process, we calculate the time evolution in the dark of the population of the spin states of the Mn initially prepared by optical pumping under CW circularly polarized excitation. The time evolution of the population of the state S$_z$=-5/2 prepared by a $\sigma+$ excitation is presented in Fig.~\ref{Fig7}(b). Without valence band mixing, $\rho_{|-5/2\rangle}$ decays with the Mn spin relaxation time $\tau_{Mn}=5 \mu s$. A decay of  $\rho_{|-5/2\rangle}$ in a few hundreds of $ns$ is obtained with a small valence band mixing term $\rho_s/\Delta_{lh}=0.05$. As illustrated in Fig.~\ref{Fig7}(b), this population decay is also slightly influenced by the coherence time of the hole-Mn system $T_2^{hMn}$.

Using in this model the valence band mixing parameter deduced from the PL spectra, the hole-Mn spin relaxation rate seems to be overestimated. However, a quantitative description is difficult with this model because of the simplified description of the coherent dynamics of the hole-Mn system and the uncertainty in the parameter $T_2^{hMn}$. In addition, we have to notice that even if we choose a gate voltage and excitation conditions for which we only observe the positively charged exciton in the PL spectra, we cannot exclude to have fluctuation of the hole in the dark. An empty dot during a fraction of the dark time would artificially increase the observed relaxation time. Further investigations are required to understand the details of this dynamics.

\section{Conclusion}

To conclude, we have shown that we could store a single hole in a singly Mn-doped II-VI semiconductor QD. The spin of the hole-Mn system can be efficiently initialized by optical pumping under quasi-resonant excitation with circularly polarized light. We demonstrated that the optical pumping of the hole-Mn system and its transverse magnetic field dependence are controlled by the electron-Mn exchange coupling within the positively charged exciton-Mn complex. Despite the magnetic anisotropy of the hole-Mn system, the relaxation time of the hole-Mn spin is shorter than a micro-second. This fast spin relaxation is a consequence of the valence band mixing always present in II-VI self-assembled QDs. In different QD systems with a larger hole confinement and weaker valence band mixing one could expect a spin memory controlled by the electrically tunable magnetic anisotropy of the hole-Mn complex. Nevertheless, a relaxation time in the micro-second range is sufficient to exploit these positively charged Mn-doped QDs for an optical ultrafast coherent control of the Mn spin \cite{Reiter2012,Reiter2013}. Pulsed resonant excitation could be used to create and annihilate the positively charged exciton and deterministically control the electron-Mn flip-flops to manipulate the Mn spin.

\begin{acknowledgments}

This work was realized in the framework of the CEA (INAC) / CNRS (Institut N\'eel) joint research team "NanoPhysique et SemiConducteurs".

\end{acknowledgments}


\begin{thebibliography}{}

\bibitem{Gambardella2003} P. Gambardella, S. Rusponi, M. Veronese, S.S. Dhesi, C. Grazioli, A. Dallmeyer, I. Cabria, R. Zeller, P.H. Dederich, K. Kern, C. Carbone, H. Brune, Science {\bf 300}, 1130 (2003).
\bibitem{Oberg2014} J. C. Oberg, M. Reyes Calvo, F. Delgado, M. Moro-Lagares, D. Serrate, D. Jacob, J. Fernandez-Rossier, C. F. Hirjibehedin, Nature Nano. {\bf 9}, 64 (2014) and references therein.
\bibitem{Bogani2008} L. Bogani, W. Wernsdorfer, Nat. Mater. {\bf 7}, 179 (2008) and references therein.

\bibitem{Besombes2004} L. Besombes, Y. Leger, L. Maingault, D. Ferrand, H. Mariette and J. Cibert, Phys. Rev. Lett. {\bf 93}, 207403 (2004).
\bibitem{Kudelski2007} A. Kudelski, A. Lemaitre, A. Miard, P. Voisin, T.C.M. Graham, R.J. Warburton and O. Krebs, Phys. Rev. Lett. {\bf 99}, 247209 (2007).
\bibitem{Kobak2014} J. Kobak, T. Smolenski, M. Goryca, M. Papaj, K. Gietka, A. Bogucki, M. Koperski, J.-G. Rousset, J. Suffczynski, E. Janik, M. Nawrocki, A. Golnik, P. Kossacki, W. Pacuski, Nature Com. {\bf 5}, 3191 (2014).
\bibitem{Besombes2012} L. Besombes, C.L. Cao, S. Jamet, H. Boukari, J. Fernandez-Rossier, Phys. Rev. B  {\bf 86}, 165306 (2012).
\bibitem{Krebs2013} O. Krebs, A. Lemaitre, Phys. Rev. Lett. {\bf 111}, 187401 (2013).

\bibitem{Qazzaz1995} M. Qazzaz, G. Yang, S.H. Xin, L. Montes, H. Luo, J.K. Furdyna, Solid State Communications {\bf 96}, 405 (1995).
\bibitem{Causa1980} M.T. Causa, M. Tovar, S.B. Oseroff, R. Calvo, W. Giriat, Phys. Lett. {\bf A77}, 473 (1980).
\bibitem{LeGall2009} C. Le Gall, L. Besombes, H. Boukari, R. Kolodka, J. Cibert, H. Mariette, Phys. Rev. Lett. {\bf 102}, 127402 (2009).
\bibitem{LeGall2010} C. Le Gall, R. S. Kolodka, C. L. Cao, H. Boukari, H. Mariette, J. Fernandez-Rossier, L. Besombes, Phys. Rev. B {\bf 81}, 245315 (2010).

\bibitem{Vyborny2012} K. Vyborny, J. E. Han, R. Oszwaldowski, I. Zutic, A. G. Petukhov, Phys. Rev. B {\bf 85}, 155312 (2012).

\bibitem{LeGall2011} C. Le Gall, A. Brunetti, H. Boukari, L. Besombes, Phys. Rev. Lett. {\bf 107}, 057401 (2011).
\bibitem{Jamet2013} S. Jamet, H. Boukari, L. Besombes, Phys. Rev. B {\bf 87}, 245306 (2013).
\bibitem{Besombes2014} L. Besombes, H. Boukari, Phys. Rev. B {\bf 89}, 085315 (2014).
\bibitem{Leger2006} Y. Leger, L. Besombes, J. Fernandez-Rossier, L. Maingault, H. Mariette, Phys. Rev. Lett. {\bf 97}, 107401 (2006).

\bibitem{Fernandez2006} J. Fernandez-Rossier, Phys. Rev. B {\bf 73}, 045301 (2006).
\bibitem{Leger2007} Y. Leger, L. Besombes, L. Maingault, H. Mariette, Phys. Rev. B {\bf 76}, 045331 (2007).
\bibitem{Besombes2005} L. Besombes, Y. Leger, L. Maingault, D. Ferrand, H. Mariette, J. Cibert, Phys. Rev. B {\bf 71}, 161307(R) (2005).
\bibitem{Trojnar2013} A. H. Trojnar, M. Korkusinski, U. C. Mendes, M. Goryca, M. Koperski, T. Smolenski, P. Kossacki, P. Wojnar, P. Hawrylak, Phys. Rev. B {\bf 87}, 205311 (2013).

\bibitem{Govorov2005} A. O. Govorov, A. V. Kalameitsev, Phys. Rev. B {\bf 71}, 035338 (2005).
\bibitem{Baudin2011} E. Baudin, E. Benjamin, A. Lemaitre, O. Krebs, Phys. Rev. Lett. {\bf 107}, 197402 (2011).
\bibitem{Goryca2009} M. Goryca, T. Kazimierczuk, M. Nawrocki, A. Golnik, J. A. Gaj, P. Kossacki, P. Wojnar, G. Karczewski, Phys. Rev. Lett. {\bf 103}, 087401 (2009).
\bibitem{Glazov2007} M.M. Glazov, E.L. Ivchenko, L. Besombes, Y. Leger, L. Maingault, H. Mariette, Phys. Rev. B {\bf 75}, 205313 (2007).

\bibitem{LeGall2012} C. Le Gall, A. Brunetti, H. Boukari, L. Besombes, Phys. Rev. B {\bf 85}, 195312 (2012).
\bibitem{Woods2004} L. M. Woods, T.L. Reinecke, R. Kotlyar, Phys. Rev. B {\bf 69}, 125330 (2004).
\bibitem{Roszak2007} K. Roszak, V.M. Axt, T. Kuhn, P. Machnikowski, Phys. Rev. B {\bf 76}, 195324 (2007).
\bibitem{Cao2011} C.L. Cao, L. Besombes, J. Fernandez-Rossier, Phys. Rev. B {\bf 84}, 205305 (2011).
\bibitem{Bounouar2012} S. Bounouar, C. Morchutt, M. Elouneg-Jamroz, L. Besombes, R. Andre, E. Bellet-Amalric, C. Bougerol, M. Den Hertog, K. Kheng, S. Tatarenko, J. Ph. Poizat, Phys. Rev. B {\bf 85}, 035428 (2012).

\bibitem{Exter2009} M.P. van Exter, J. Gudat, G. Nienhuis, D. Bouwmeester, Phys. Rev. A {\bf 80}, 023812 (2009).
\bibitem{Roy2011} C. Roy, S. Hughes, Phys. Rev. X {\bf 1}, 021009 (2011).

\bibitem{Besombes2008} L. Besombes, Y. Leger, J. Bernos, H. Boukari, H. Mariette, J.P. Poizat, T. Clement, J. Fernandez-Rossier, R. Aguado, Phys. Rev. B {\bf 78} 125324 (2008).

\bibitem{Reiter2012} D.E. Reiter, T. Kuhn, V.M. Axt, Phys. Rev. B {\bf 85}, 045308 (2012).
\bibitem{Reiter2013} D.E. Reiter, V.M. Axt, T. Kuhn, Phys. Rev. B {\bf 87}, 115430 (2013).

\end{thebibliography}
\end{document}